\documentclass[usegraphicx]{mn2e}
\usepackage{graphicx}
\begin{document}
\title[Ionized accretion disc models of PG 1404+226]{Investigating ionized disc models of the variable narrow-line Seyfert 1 PG 1404+226}
   \author[J. Crummy et al.]{J. Crummy$^1$\thanks{E-mail: jc@ast.cam.ac.uk}, A.C. Fabian$^1$, W.N. Brandt$^2$, and Th. Boller$^3$\\
   $^1$Institute of Astronomy, Madingley Road, Cambridge, CB3 0HA\\
   $^2$Department of Astronomy and Astrophysics, The Pennsylvania State University, 525 Davey Lab, University Park, PA 16802, USA\\
   $^3$Max-Planck-Institut f\"{u}r extraterrestrische Physik, Postfach 1312, 85741 Garching, Germany}

\maketitle

\begin{abstract}
We investigate the use of relativistically blurred photoionized disc models on an \textit{XMM-Newton} observation of the Narrow Line Seyfert 1 galaxy PG 1404+226. The model is designed to reproduce the radiation from the inner accretion disc around a Kerr black hole, and is more successful at fitting the spectrum than models based on a thermal soft excess. The source varies strongly over the course of the observation, and the disc model works over all observed flux states. We conclude that it is a useful tool in the study of certain quasars.
\end{abstract}

% keywords for MNRAS etc.

\section{Introduction}
\begin{figure}
\begin{center}\includegraphics[angle=270, width=0.45\textwidth]{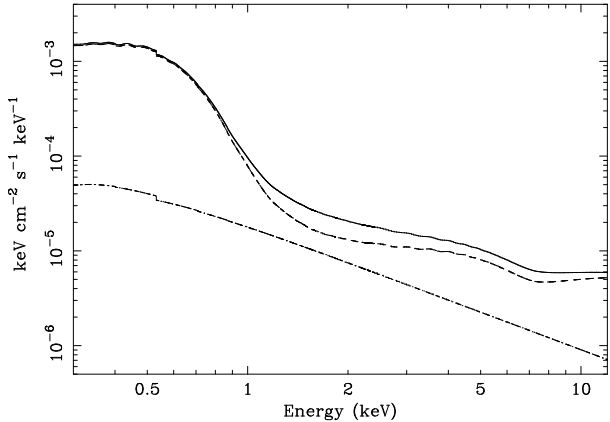}
\caption{The ionized disc model as fit to the complete observation (see the top of Table \ref{pg1404_reflion_table} for the details of the fit). The dashed line is the power-law component, the dotted line is the reflection component, and the solid line is their total.}
\label{pg1404_folded}
\end{center}
\end{figure}

Narrow Line Seyfert 1 (NLS1) galaxies are a class of Active Galactic Nuclei (AGN) first studied by Osterbrock \& Pogge (1985). The class is defined by its optical properties: the Balmer lines such as H$\beta$ have a FWHM $<$ 2000~km/s, the ratio of the \textsc{[O\thinspace III]} $\lambda$5007~\AA\ forbidden line to the H$\beta$ lines is $<$ 3, and emission lines from Fe\thinspace \textsc{II} or higher ionization states are often present. These properties are discussed in, e.g. Boroson \& Green (1992). In the X-ray band Seyfert galaxies are usually well described by a power-law and a `soft excess', more emission below $\sim$1~keV than expected from an extrapolation of the power-law spectrum observed at higher energies. Modelling the soft excess as a black-body usually results in a good fit. The soft excess is notable for having similar temperatures over a wide range of objects, and is the subject of some debate: if it is thermal in origin it may be due to a slim disc (e.g  Abramowicz et al. 1988), Gierli\'{n}ski \& Done (2004) show that soft excess temperature in a sample of radio-quiet PG quasars is 0.1 --\ 0.2~keV over a large range of luminosity, and put forward the idea that the soft excess is an illusion caused by relativistically blurred strong absorption. This paper is based on the idea that it is due to photoionized emission blurred by relativistic motion in an accretion disc.\\
Seyferts are highly variable, in the optical on time-scales of weeks to years, and in the X-ray on time-scales of hours or even less. NLS1s are more variable in the X-ray band than typical Seyfert 1s, implying they are more compact (Leighly 1999a). X-rays are generated only in a small energetic area of an AGN, i.e. near the central engine. The detection of iron lines broadened by relativistic effects in some sources shows that the central energy source is matter accreting onto a black hole (e.g. Tanaka 1995; Vaughan \& Fabian 2004).\\
This concept is extended in relativistically blurred  photoionized disc models, the most recent of which (Ross \& Fabian 2005) is used in this paper. This model is intended to simulate radiation from the inner regions of a black hole accretion disc, and it includes atomic physics and general relativistic effects. The underlying spectrum is due to a slab of optically thick gas of constant density illuminated by a power-law, which produces fluorescence lines with a continuum of reflected photons. A power-law with the spectral index of the illuminating photons is added to include light from the illuminating object (e.g. from the base of a jet or a hot corona). The power-law component dominates the observed spectrum in many NLS1s but in PG 1404+226 the soft excess is the most powerful component (Fig. \ref{pg1404_folded}). This spectrum is then relativistically blurred as calculated for an accretion disc around a Kerr black hole. The Laor (1991) line profile (standard in \textsc{xspec}) is used as a convolution kernel to do this.\\
This model is more physically motivated than the ad hoc models usually adopted (i.e. a power-law and a black-body to model the soft excess). We show that this new model is a better fit to the source than previous models and adequately accounts for variability.\\

\section{The source}
PG 1404+226 is a Narrow Line Seyfert 1 galaxy, with a redshift of 0.098. Wang \& Lu (2001), using the relation with \textsc{[O\thinspace III]} width, find the central black hole mass to be $1.0\times 10^{7}$~M$_{\odot}$, with an error of $\sim$0.5~dex. During the observation we observe a flux of $1.1 \times 10^{-12}$~ergs cm$^{-2}$ s$^{-1}$ in the 0.3 -- 12.0~keV range, which we convert (using the inverse square law with a Hubble parameter of 70~km s$^{-1}$ Mpc$^{-1}$) to a luminosity of $2.3 \times 10^{43}$~erg s$^{-1}$.\\
The source has previously been observed with the \textit{ASCA} satellite (e.g. Vaughan et al. 1999), and is included in the two-part review by Leighly (1999a,b). The \textit{ASCA} observation showed PG 1404+226 to be strongly variable, with a clear soft excess. The source showed possible apparent absorption around 1~keV, which is discussed in detail in e.g. Leighly et al. (1997). PG 1404+226 did not show an iron line. As this article will show, the 1~keV feature is a natural consequence of an ionized disc.\\

\section{Observations and data reduction}\label{reduction}
PG 1404+226 was observed on-axis with \textit{XMM-Newton} during revolution 0279 (2001-06-17 -- 2001-06-18). All \texttt{EPIC} cameras were operated in \texttt{PrimeFullWindow} mode with the filter \texttt{Thin1}. The \texttt{RGS} instruments were used in the \texttt{HighEventRateWithSES} mode. The \texttt{OM} did not take data. During the observation the \texttt{EPIC MOS} took 20.8~ks of data, the \texttt{EPIC pn} 18.3~ks and the \texttt{RGS} instruments 21.4~ks.\\
The Observation Data Files (ODFs) were reduced in the standard way using \textsc{sas 5.4.1} to produce event lists. Light curves and spectra were created for the sources by using a circular, source-centred extraction region of 20 arcsec radius for the \texttt{pn} and 30 arcsec for the \texttt{MOS}. The smaller than standard extraction radii were chosen to exclude as much as possible of the strong background. The background light curves and spectra were extracted using regions away from any sources and chip-gaps. The presence of strong background flaring in the observation was noted and compensated for using a Good Time Interval (GTI) file, leaving 16.9~ks for the \texttt{pn} and 20.1~ks for the \texttt{MOS}.\\
The \texttt{pn} images were also examined for Out-Of-Time (OOT) events and pile-up. Any OOT events in PG 1404+226 are unobservable due to a low signal to noise ratio, so no OOT correction was used. Use of the \textsc{sas} task \textsc{epatplot} showed no pile-up.\\
The \texttt{MOS} cameras are not accurately calibrated against the \texttt{pn} (Kirsch et al. 2004), and the \texttt{pn} has a larger effective area and spectral range. Therefore all fits were done on \texttt{pn} data alone, and checked against the \texttt{MOS} for consistency. The \texttt{MOS} gives a flatter spectral index than the \texttt{pn}, but no other serious differences were noted. The \texttt{RGS} data were reduced and found to be consistent with the other instruments, however, the statistics are dominated by the \texttt{pn} and \texttt{MOS} so \texttt{RGS} data were not included in any of the fits. The limits in accurate calibration were taken as 0.3~keV and 12.0~keV for the \texttt{pn} cameras, and the \texttt{MOS} was used over the 0.3~keV -- 10.0~keV range. The data reduction yielded 7807 (\texttt{pn}) source photons (including an estimated 232 background photons).\\
The spectral analysis was performed using spectra grouped so that each bin contained at least 20 source counts, to ensure that $\chi^2$ statistics would give reasonable results. Response files generated by \textsc{sas} were used, and the spectra were examined using \textsc{xspec v11.3} (Arnaud 1996). A Galactic hydrogen absorption column of $2.14 \times 10^{20}$~cm$^{-2}$ was taken from the \textsc{nh} ftool (Dickey \& Lockman 1990). Fits were also performed allowing for free hydrogen columns at the source redshift, no evidence for any extra absorption was found. All quoted errors are 90\% limits on one parameter.\\

\section{Analysis}\label{pg1404}
\begin{figure}
\begin{center}\includegraphics[angle=270, width=0.45\textwidth]{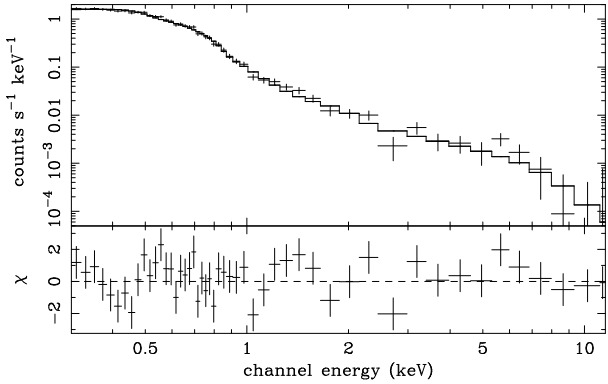}
\caption{The \texttt{pn} spectrum of PG 1404+226, showing data (crosses) and the ionized disc model fit (line). The lower panel shows the residuals to the model, with error bars scaled to a size of one. The data has been re-binned for display.}
\label{pg1404_spectrum}
\end{center}
\end{figure}

The data were reduced and spectra were created as described in Section \ref{reduction}. The spectrum is shown in Fig. \ref{pg1404_spectrum}. We fit the data with the standard black-body and power-law. Adding a red-shifted edge fixed at a source-frame energy of 0.87~keV (the \textsc{O\thinspace VIII} K absorption edge) improves the fit ($\Delta \chi^{2}$ of 18 for 1 degree of freedom). Adding a second edge at 0.74~keV (\textsc{O\thinspace VII}) does not improve the fit. Allowing the energy of the first edge to vary improves $\chi^{2}$ by 4 for 1 degree of freedom. The results are given in Table \ref{pg1404_simple_table}. To model cold iron reflection from distant gas (e.g. a torus) we added a narrow red-shifted Gaussian at 6.4~keV (in the source frame). This improves the fit further, to a $\chi^{2}_{\nu}$ of 0.916 for 190 degrees of freedom (with all other parameters consistent with those reported for the previous fit); allowing the width or energy of the line to vary does not improve the fit. An F-test gives an 89\% probability that the addition of the line is valid. Including \texttt{MOS} data does not increase the significance, an F-test gives an 84\% probability that the line is valid when all three \texttt{EPIC} observations are fit simultaneously. The use of the F-test on hypotheses on the boundary of the parameter space, such as testing for the existence of an emission line, is of questionable statistical value (Protassov et al. 2002). We therefore calibrate the F-test using simulated observations (from the \textsc{xspec} `fakeit' command), by generating data from a model without an iron line. In this data we measure an iron line at the same or higher significance (measured by an F-test) as detected in our real data 12\% of the time, therefore we conclude that the F-test is reasonably calibrated in this case and that the line is 88\% significant. The statistics are poor at these energies, so the equivalent width of the line is barely constrained; \textsc{xspec} reports the equivalent width of the line as $0.5^{+1.6}_{-0.5}$~keV. Given that the line is just on the boundary of 90\% significance, we choose not to include the line in any of our further fits. We therefore consider the free edge energy model to be our best simple fit, with a $\chi^{2}_{\nu}$ of 0.934 for 191 degrees of freedom.\\

\begin{table*}
\caption{PG 1404+226 simple model fits. $kT$ is the black-body temperature in keV, $\Gamma$ is the spectral index of the power law component, and $\tau$ is the optical depth of the absorption edge.}
\label{pg1404_simple_table}
\centering
\begin{tabular}{lcccccc}
\hline
Model & $kT$ (keV) & $\Gamma$ & Edge energy (keV) & $\tau$ & $\chi^{2}_{\nu}$ (d.o.f.) \\
\hline
power-law + black-body & $0.1109^{+0.0018}_{-0.0018}$ & $1.53^{+0.20}_{-0.14}$ & - & - & $1.04 (193)$ \\
power-law + black-body + edge & $0.1153^{+0.0037}_{-0.0021}$ & $1.67^{+0.34}_{-0.31} $ & 0.87${}^{a}$ &  $0.42^{+0.20}_{-0.16}$ & $0.950 (192)$ \\
power-law + black-body + edge & $0.1144^{+0.0019}_{-0.0019}$ & $1.81^{+0.19}_{-0.14}$ & $0.94^{+0.03}_{-0.23}$ & $0.52^{+0.14}_{-0.13}$ & $ 0.934 (191)$ \\
\hline
\end{tabular}
\\
${}^{a}$ Edge energy fixed.
\end{table*}

We also fit PG 1404+226 using the ionized disc model. The free parameters of the model are the iron abundance (Fe, in units of solar metallicity), ionization parameter ($\xi$, the number of incident ionizing photons~cm$^{-2}$s$^{-1}$), and the spectral index of the illuminating continuum ($\Gamma$, this is also the index of the power-law component we add). The spectrum is blurred with a Laor line profile, which is standard in \textsc{xspec}, and has as free parameters the emissivity index of the disc (Index), the inclination of the disc to the line of sight ($\theta$, in degrees), the inner radius of the disc ($R_{in}$, in units of gravitational radii) and outer radius (which we fix at 100 gravitational radii as it is not strongly constrained). This is intended to reproduce the spectrum from a photoionized disc around a maximally-rotating (Kerr) black hole. Fits made using relativistic blurring from a non-rotating (Schwarzschild) black hole were not found to be satisfactory; there were strong residuals below 1~keV where the blurring is insufficient to smooth out the emission and absorption features. A more rapidly rotating black hole allows for stronger blurring.\\
The parameters obtained in fitting PG 1404+226 with the ionized disc model are given in Table \ref{pg1404_reflion_table}, and the model is plotted showing the two components in Fig. \ref{pg1404_folded}. We find no evidence for absorption edges with this model. Our best ionized disc fit has a $\chi^{2}_{\nu}$ of 0.892 for 189 degrees of freedom.\\

\begin{table*}
\caption{PG 1404+226 ionized disc model fits, including spectral variation with time. See Section \ref{pg1404} for an explanation of parameters.}
\label{pg1404_reflion_table}
\centering
\begin{tabular}{lccccccc}
\hline
Time (ks) & Index & $R_{in}$ & $\theta$ & Fe & $\Gamma$ & $\xi$ & $\chi^{2}_{\nu}$ (d.o.f.) \\
\hline
0 -- 18${}^{a}$ & $ 8.9 ^{+1.1}_{-4.1} $ & $ 1.235^{+0.362}_{-0.0} $ & $58^{+7}_{-24} $ & $5.4 ^{+2.2}_{-2.8} $ & $ 2.317 ^{+0.098}_{-0.123} $ & $ 870^{+430}_{-580} $ & $ 0.892 (189) $   \\
\hline
0 -- 7.5${}^{b}$ & $ 10.0^{+0.0}_{-3.0}  $ & $    1.237^{+0.124}_{-0.002}  $ & $  67.3^{+6.6}_{-3.8}   $ & $ 2.2^{+1.2}_{-0.60}  $ & $  2.40^{+0.07}_{-0.25}  $ & $ 355^{+38}_{-167}  $ & $  0.991  (95) $ \\
7.5 -- 12 & $ 2.98^{+0.49}_{-0.84}  $ & $  1.235^{+0.772}_{-0.0}  $ & $  89.0^{+1.0}_{-4.4} $ & $ 1.8^{+8.2}_{-1.6}  $ & $ 2.808^{+0.192}_{-0.180} $ & $  16.8^{+17.3}_{-6.0}     $ & $  0.889 (92) $ \\
12 -- 14 & $7.92^{+0.89}_{-1.62}  $ & $	   1.235^{+0.332}_{-0.0}  $ & $  59.2^{+10.0}_{-2.0} $ & $ 6.0^{+3.3}_{-2.7} $ & $ 2.419^{+0.053}_{-0.209} $ & $ 510^{+2640}_{-170}  $ & $ 1.12  (68) $ \\
14 -- 18 & $ 9.07^{+0.63}_{-8.73}   $ & $    1.237^{+0.157}_{-0.002} $ & $  63.2^{+4.4}_{-18.3}  $ & $ 2.11^{+1.8}_{-0.87}  $ & $ 3.00^{+0.0}_{-0.29}  $ & $  302.4^{+9.2}_{-7.9}   $ & $    0.983 (96) $ \\
\hline
0 -- 7.5${}^{c}$ & $9.91^{+0.09}_{-1.64} $ & $  1.235^{+0.065}_{-0.0} $ & $	  61.8^{+3.6}_{-10.6} $ & $  6.0^{+1.7}_{-1.7} $ & $  2.401^{+0.102}_{-0.032} $ & $  610^{+580}_{-220}$ & $ 0.967 (338) $ \\
7.5 -- 12 &\\
12 -- 14 &\\
14 -- 18 &\\
\hline
0 -- 7.5${}^{d}$ &  $9.74^{+0.26}_{-3.70}$ & $ 1.235^{+0.218}_{-0.0}$ & $  56.6^{+9.2}_{-12.2}$ & $ 6.1^{+2.3}_{-1.7}$ & $ 2.163^{+0.197}_{-0.042}$ & $ 710^{+600}_{-270} $ & $ 0.918 (335) $ \\
7.5 -- 12 & & & & & $ 2.383^{+0.106}_{-0.051} $ & & \\
12 -- 14 & & & & & $ 2.332^{+0.102}_{-0.107} $ & & \\
14 -- 18 & & & & & $ 2.400^{+0.095}_{-0.053} $ & & \\
\hline
\end{tabular}
\\
${}^{a}$ The complete observation. ${}^{b}$ All time periods fit independently. ${}^{c}$ All time periods fit simultaneously, with all parameters except the normalizations of the reflection and power-law components fixed. ${}^{d}$ As ${}^{c}$, but with free spectral index.
\end{table*}

\subsection{Variability}
\begin{figure}
\begin{center}
\includegraphics[angle=270, width=0.45\textwidth]{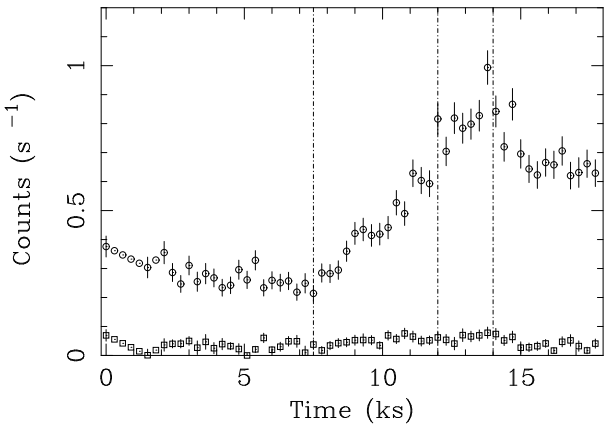}
\caption{The background subtracted light-curve of PG 1404+226, taken from the \texttt{pn} camera. The circles show soft counts from 0.3 -- 1.0~keV, and the squares hard counts from 1.0 -- 8.0~keV. The errors are Poisson, and points without associated errors have been interpolated over to avoid periods of background flaring. The vertical lines show the boundaries between the time periods analysed in Tables~\ref{pg1404_reflion_table} and \ref{pg1404_vary_table}.}
\label{pg1404_lightcurve}
\end{center}
\end{figure}

PG 1404+226 varied by a factor of $\sim$3 over the course of the observation, as shown in  Fig. \ref{pg1404_lightcurve}. The Figure shows soft (0.3 -- 1.0~keV) and hard (1.0 -- 8.0~keV) count rates. Counts above 8~keV are not included as the source is extremely weak at those energies. The hard count rate does not strongly vary whereas the soft count rate increases during the observation; this indicates that the spectral shape changes. To investigate this change we divide the observation into four different periods suggested by the shape of the light-curve, as indicated by the vertical lines on the Figure. We fit these using our best simple model, testing to see if adding the absorption edge (at a fixed energy of 0.9~keV in the source frame) improves the fit. The results are given in Table \ref{pg1404_vary_table}. Adding an edge results in an improvement to the fit in all time periods but the first one. If an edge is included in the first period fit \textsc{xspec} reports an optical depth of $0.05^{+0.41}_{-0.05}$, with a negligible change in  $\chi^{2}$ for the loss of a degree of freedom. To further investigate this we perform simulated observations to evaluate whether we would detect an absorption edge with a $\tau$ of 0.9. We measure an optical depth $>$ 0.05 in $\sim$99.9\% of our simulated observations, therefore we are confident that the absence of an absorption edge in the first time period is not due to insufficient data quality.\\
\begin{table*}
\caption{PG 1404+226 investigation of spectral variation over time. $kT$, $\Gamma$ and $\tau$ as in Table \ref{pg1404_simple_table}. `Time' gives the start and end times (in ks, counting from the start time) of the section of the observation being fit, as marked on the light-curve given in Fig. \ref{pg1404_lightcurve}.}
\label{pg1404_vary_table}
\centering
\begin{tabular}{lccccc}
\hline
Time (ks) & $kT$ (keV) & $\Gamma$ & Edge energy (keV) & $\tau$ & $\chi^{2}_{\nu}$ (d.o.f.) \\
\hline
0 -- 7.5${}^{a}$ & $0.1189^{+0.0048}_{-0.0050}$ &  $1.55^{+0.65}_{-0.33} $ & $ 0.9{}^{b}$ & $0.0{}^{c} $ & $ 0.987 (99) $ \\
7.5 -- 12 & $ 0.1175^{+0.0054}_{-0.0052} $ & $  1.67^{+0.33}_{-0.22} $ & $ 0.9{}^{b} $ & $  0.33^{+0.36}_{-0.33} $ & $  0.897   (95) $ \\
12 -- 14 & $ 0.1158^{+0.0058}_{-0.0054} $ & $  1.74^{+0.89}_{-0.79}$ & $ 0.9{}^{b} $ & $  0.77^{+0.53}_{-0.44}	$ & $  1.14 (71) $ \\
14 -- 18 & $ 0.1103^{+0.0038}_{-0.0038}$ & $   2.02^{+0.80}_{-0.64}$ & $ 0.9{}^{b} $ & $  0.88^{+0.39}_{-0.36} $ & $ 0.980   (99) $ \\
\hline
0 -- 7.5${}^{d}$ & $ 0.1136^{+0.0018}_{-0.0024} $ & $ 1.854^{+0.168}_{-0.148} $ & $ 0.906^{+1.001}_{-0.030} $ & $  0.0^{+0.149}_{-0.0}  $ & $ 0.971 (337) $ \\ 
7.5 -- 12 & & & & $ 0.938^{+0.401}_{-0.271} $ & \\
12 -- 14 & & & & $  0.645^{+0.407}_{-0.240} $ & \\
14 -- 18 & & & & $  0.986^{+0.351}_{-0.220} $ & \\
\hline
\end{tabular}
\\
${}^{a}$ All parameters independent. ${}^{b}$ Edge energy fixed. ${}^{c}$ Adding an edge does not improve the fit, so $\tau$ was fixed at zero. ${}^{d}$ Investigating variation in $\tau$; $kT$, $\Gamma$ and Edge Energy are fit simultaneously.
\end{table*}

We also fit the same time periods using the ionized disc model, with the results in Table \ref{pg1404_reflion_table}. The ionized disc model and simple model fit the data about equally well when analysing the shorter time periods individually.\\
It has been suggested that the variability in some sources is due to changes in the normalization of the underlying continuum, with a fairly constant reflection component from an ionized disc (Fabian et al. 2002, Vaughan \& Fabian 2004). To investigate this we simultaneously fit all four time periods, allowing each to have independent reflection and power-law component normalizations and tying all the other parameters together. We investigate possible changes in spectral index of the underlying continuum by repeating the previous fit with a free spectral index ($\Gamma$). The results from both of these fits are reported in Table \ref{pg1404_reflion_table}. We further attempt to fit the data with no power-law component at all. This did not improve the fits so we do not report the results.\\
We similarly fit the four periods simultaneously with the simple model, tying the black-body temperature, power-law index and edge energy together but allowing the normalizations and optical depths to vary independently. The results are given in Table \ref{pg1404_vary_table}. Our best fit to the data is with the ionized disc model, simultaneously fitting all parameters except the normalizations and $\Gamma$; this has a $\chi^{2}_{\nu}$ of 0.918 for 335 degrees of freedom.\\

\section{Discussion}
\subsection{Simple model}
The simple model is an acceptable fit to the data. This model includes absorption at $\sim$0.9~keV, which is likely to be due to the \textsc{O\thinspace VIII} K absorption edge at 0.87~keV. The $> 0.87$~keV energy might imply the absorption originates in a slightly relativistic outflow, although the data do not constrain this (see Leighly et al. 1997).\\
We learn more when we consider spectral variability over the observation. The absorption appears to `switch on' some way into the observation, as shown in Table \ref{pg1404_vary_table}. The absorption increases at the same time as the luminosity of the source increases (at $\sim$7.5~ks, see Fig. \ref{pg1404_lightcurve}), and we have eliminated the possibility that this is a statistical effect.  If the absorption is due to \textsc{O\thinspace VIII}, the lack of strong \textsc{O\thinspace VII} absorption implies the absorbing gas has a high ionization parameter; this means it is close to the central black hole and the rapid change implied by the variability of the feature is plausible. This variation in absorption has been previously investigated by Dasgupta et al. (2004), who understand it within the model of absorption due to a disc wind - as the source brightens more matter is driven off in the wind and the absorption intensifies. We have shown that the ionized disc model naturally explains this apparent increase in absorption.\\

\subsection{Ionized disc model}
The ionized disc model is the best fit to the entire observation. This model has the benefit of being self-consistent and not ad hoc. It does not require any additional absorption features to give a good fit, the model itself reproduces the dip in the spectrum. When the spectral variability of the source is investigated using the ionized disc model, it accounts for it naturally; the spectral index of the illuminating powerlaw varies over the course of the observation. This fit (at the bottom of Table \ref{pg1404_reflion_table}) has a $\chi^{2}$ 20 lower than the equivalent simple fit (Table \ref{pg1404_vary_table}) for two fewer degrees of freedom. We conclude that the ionized disc model is the most satisfactory explanation of the data.\\
The fit parameters give us information about the central source. Firstly, it is evident from Table \ref{pg1404_reflion_table} that the inclination of the source is quite high; around $60^{\circ}$. This does not fit very well with the unified model of AGN (Antonucci \& Miller 1985, Antonucci 1993), in which Seyfert 1 galaxies are those visible at low inclinations, with no molecular torus in the line of sight. However, PG 1404+226 is fairly luminous, and there have been suggestions that brighter quasars tend to show less absorption from a torus (Lawrence 1991, Brandt \& Hasinger 2005). The high inclination also implies that the brightness of the source is partly due to relativistic Doppler beaming from the rotation of the disc. We investigated the effect of inclination angle on flux within our model, and find that the brightness is decreased by a factor of $\sim$5 if the inclination is changed to 0$^{\circ}$ (all other parameters unchanged), and increased by a factor of $\sim$3 at 90$^{\circ}$. The measured luminosity is therefore a reasonable order-of-magnitude estimate to the intrinsic luminosity of the source.\\
Information on the rotation can also be obtained. Note that fits were performed assuming a Kerr black hole, and that Schwarzschild fits were unacceptable (Section \ref{pg1404}). The measured inner disc radius provides further evidence that the black hole is rotating - it is at the last stable orbit radius for a maximally rotating black hole; well inside the plunging region for a non-rotating black hole. The black hole rotation is not a free parameter in this analysis so we must be careful in drawing conclusions, however it seems that a rapidly rotating black hole is likely. It is also a possibility that radiation from the plunging region is important (see Krolik \& Hawley 2002).\\
The power-law component of this source is weak compared to the reflection component, i.e. the reflection fraction is very high. This may be due to light bending effects on the primary continuum. If the continuum source is compact and close to the black hole almost all the radiation it emits is bent onto the disc rather than escaping to the observer; this reduces the observed power-law flux while enhancing the disc illumination, and therefore the reflection component (Miniutti et al. 2003; Miniutti \& Fabian 2004). To produce a reflection dominated spectrum in this model the primary source in PG 1404+226 must be within 3 -- 4 gravitational radii of the black hole. Reflected radiation escapes the black hole more easily than the illuminating continuum as it is emitted by rapidly moving matter and tends to be beamed along the plane of the disc to the observer. The high inclination therefore also contributes to the high reflection fraction.\\
This model raises the question of how energy is transported through the system. The original source of energy is the matter falling into the gravitational potential of the central black hole. This heats the electrons that provide the illuminating continuum, and the energy reflects off the disc and escapes. The question of how the in-falling matter heats the source of the illuminating continuum is an unsolved one, magnetic reconnection is a possibility, as is emission from the base of a weak jet, or shocks in a failed jet (Ghisellini, Haardt \& Matt 2004). In the case of a jet the spin of the black hole may also provide energy.\\
Given the small disc inner radius we calculate for the source, most of the X-ray emission is coming from within a few gravitational radii of the black hole, so strong gravity effects are expected. This small size also explains the rapid variability. If the illuminating continuum is, e.g. UV radiation being Compton up-scattered by hot electrons, the change in spectral index could be due to a change in temperature of the electrons, and the extra flux could be due to an increase in emitting area.\\

\section{Conclusions}
We have shown that:
\begin{itemize}
\item Physically plausible models based on relativistically blurred reflection from photoionized discs fit observations to PG 1404+226 better than the typical ad hoc simple model of a black-body and power-law. In particular, a `soft excess' at low X-ray energies is a natural consequence of a blurred ionized disc; the `power-law' is largely made up of broadened iron line emission and the `soft excess' of other blurred lines, plus the Compton reflection component. The ionized disc model also leads naturally to apparent absorption features at low X-ray energies.
\item The X-ray emitting accretion disc in PG 1404+226 has an inclination of $\sim60^{\circ}$.
\item The central black hole is most likely strongly rotating, alternatively, radiation from the plunging region may be important.
\item The source has strong and complex spectral variability, which is well described within the ionized disc model by a variation in the spectral index of the illuminating continuum, as well as variation in the flux of the two components.
\item If the alternative simple model is adopted, the spectral variation is reasonably fit with a rapid increase in absorption at $\sim$0.9~keV coincident with the increase in luminosity of the source, plus some variation in the other model parameters.
\end{itemize}
The new relativistically blurred photoionized disc model of Ross \& Fabian (2005) is clearly a valuable tool in investigating NLS1s, and in sources where an ionized disc fit is appropriate information about the central conditions can be obtained. The application of these models to more, brighter sources (currently in progress) should lead to advances in our knowledge of AGN.

\section*{Acknowledgments}
J.C. is a PPARC funded PhD student. A.C.F. thanks the Royal Society for support. W.N.B. acknowledges support from NASA grant NAG5-9924 and NASA LTSA grant NAG5-13035. The \textit{XMM-Newton} satellite is an ESA science mission (with instruments and contributions from NASA and ESA member states). This work made use of the NASA/IPAC Extragalactic Database. J.C. would like to thank Luigi Gallo and Giovanni Miniutti for many helpful comments, and Randy Ross, Simon Vaughan, and Roderick Johnstone for contributing \textsc{xspec} models and code.\\

\section*{References}
	
Abramowicz M. A., Czerny B., Lasota J. P., Szuszkiewicz E., 1988, ApJ, 332, 646\\
%Extreme slim discs
Antonucci R., 1993, ARA\&A, 31, 473\\
%Unified model reviewed
Antonucci R.R.J., Miller J.S., 1985, ApJ, 297, 621\\
%Unified model proposed
Arnaud K.A., 1996, ADASS 5, 17A\\
%(xspec, the first 10 years)
Boroson T.A., Green R.F., 1992, ApJS, 80, 109\\
%Emission line properties of quasars
Brandt W.N., Hasinger G., 2005, astro-ph/0501058\\
%Deep extragal surveys, annual review.
Dasgupta S., Rao, A.R., Dewangan G.C., Agrawal V.K., 2005, ApJ, 618, L87\\
%varying absorption
Dickey J.M., Lockman F.J., 1990, ARA\&A, 28, 215\\
%galactic hydrogen
Fabian A.C., et al., 2002, MNRAS, 335, L1\\
%8 co-authors
%long hard look at mcg6
Ghisellini G., Haardt F., Matt G., 2004, A\&A, 413, 535\\
% aborted jets
Gierli\'{n}ski M., Done C., 2004, MNRAS, 349, L7\\
%(is the soft excess real - absorption...)
Kirsch M.G.F., et al, 2004, Proc. SPIE, 5488, 103\\
% Kirsch and 9 co-authors.
%Xmm (cross)-calibration
Krolik J.H., Hawley J.F., 2002, ApJ, 573, 754\\
%radiation from the plunging region
Laor A., 1991, ApJ, 376, 90\\
%(Laor profile as in kdblur)
Lawrence A., 1991, MNRAS, 252, 586\\
%torus angle varies with luminosity
Leighly K., 1999a, ApJSS, 125, 297\\
%(21 ASCA quasars, observations and variability)
Leighly K., 1999b, ApJSS, 125, 317\\
%(part 2, spectral)
Leighly K., Mushotzky R., Nandra K., Forster K., 1997, ApJ, 489, L25\\
%(rel abs line 1404)
Miniutti G., Fabian A.C., 2004, MNRAS, 349, 1435\\
%(light bending model)
Miniutti G., Fabian A.C., Goyder R., Lasenby A.N., 2003, MNRAS, 344, 22\\
%(light bending in MCG-6)
Osterbrock D.E., Pogge R.W., 1985, ApJ, 297, 166\\
%(1st major NLS1 paper...)
Protassov R., van Dyk D.A., Connors A., Kashyap V.L., Siemiginowska A., 2002, 571, 545\\
%F-tests
Ross R.R., Fabian A.C., 2005, MNRAS, 358, 211\\
%(reflion model)
Tanaka Y., et al., 1995, Nature, 375, 659\\
%ASCA mcg6, line profile, 11 co-authors
Vaughan S., Fabian A.C., 2004, MNRAS, 348, 1415\\
%Look hard look at mcg6 part 2, 2 component model etc.
Vaughan S., Reeves J., Warwick R., Edelson R., 1999, 309, 113\\
%X ray spectral complexity in NLS1s
Wang J.-M., Netzer H., 2003, A\&A, 398, 927\\
%(extreme slim discs)
Wang T., Lu Y., 2001, A\&A, 377, 52\\
%(variability properties of AGN)

\end{document}